\documentclass{aa}

\usepackage{graphicx}
\usepackage{xcolor}
\usepackage[colorlinks=true,citecolor=blue]{hyperref}

\usepackage[normalem]{ulem}  

\usepackage{txfonts}
\graphicspath{ {./Pictures/} }
\begin{document}

   \title{High-contrast detection of exoplanets with a kernel-nuller at the VLTI} 

   \author{Peter Marley Chingaipe
          \inst{1},
          Frantz Martinache\inst{1},
          Nick Cvetojevic\inst{1},
          Roxanne Ligi\inst{1},
          David Mary\inst{1},
          Mamadou N'Diaye\inst{1},
          Denis Defrère\inst{2},
          Michael J. Ireland\inst{3}
          }
   \authorrunning{Peter Marley Chingaipe et al.} 

   \institute{Laboratoire Lagrange, Université Côte d’Azur,
              Observatoire de la Côte d’Azur, CNRS, Parc Valrose, Bât. H. Fizeau, 06108 Nice, France.\\
         \and
         Institute of Astronomy, KU Leuven, Celestijnenlaan 200D, 3001, Leuven, Belgium.\\
         \and
         Research School of Astronomy and Astrophysics, College of Science, Australian National University, Canberra, Australia, 2611.\\
             }

   \date{Received February 10, 2023 / Accepted April 24, 2023}

 
  \abstract
   {The conventional approach to direct imaging has been the use of a single aperture coronagraph with wavefront correction via extreme adaptive optics. Such systems are limited to observing beyond an inner working (IWA) of a few $\lambda /D$. Nulling interferometry with two or more apertures will enable detections of companions at separations at and beyond the formal diffraction limit.}
   {This paper evaluates the astrophysical potential of a kernel-nuller as the prime high-contrast imaging mode of the Very Large Telescope Interferometer (VLTI).}
   {By taking into account baseline projection effects which are induced by Earth rotation, we introduce some diversity in the response of the nuller as a function of time. This response is depicted by transmission maps. We also determine whether we can extract the astrometric parameters of a companion from the kernel outputs, which are the primary intended observable quantities of the kernel-nuller. This then leads us to comment on the characteristics of a possible observing program for the discovery of exoplanets.}
   {We present transmission maps for both the raw nuller outputs and their subsequent kernel outputs. To further examine the properties of the kernel-nuller, we introduce maps of the absolute value of the kernel output. We also identify 38 targets for the direct detection of exoplanets with a kernel-nuller at the focus of the VLTI.}
   {With continued upgrades of the VLTI infrastructure that will reduce fringe tracking residuals, a kernel-nuller would enable the detection of young giant exoplanets at separations $<$ 10 AU, where radial velocity and transit methods are more sensitive.}


   \maketitle
%

\section{Introduction}

   The use of nulling interferometry as a way of detecting and characterising exoplanets has experienced a resurgence in recent years owing to new innovative concepts and advances in technology. Enabling the implementation of nulling architectures are mature infrastructure for long-baseline interferometry. These include the Very Large Telescope Interferometer (VLTI) and the Centre for High Angular Resolution Astronomy (CHARA), which now combine 4/6 telescopes routinely. Current active endeavours in nulling interferometry for extrasolar planets are the NOTT project (formerly Hi-5, \cite{2018ExA....46..475D}) and LIFE \citep{2019EPSC...13..327Q}. 
   \smallbreak
   Towards the end of the 70's, \cite{1979Icar...38..136B} introduced the concept of a two-telescope space-based nulling interferometer, aptly known as the Bracewell interferometer. Later on \cite{1997ApJ...475..373A} combined two Bracewell interferometers thus achieving greater suppression of the on-axis source and a superior $(u,v)$ coverage. Also drawing from Bracewell's idea, \cite{2013PASP..125..951G} improved the attenuation of the on-axis source by optimising the recombination process of a nuller. This recombination process serves to reject the light from an on-axis source (normally a star) by means of coherent destructive interference between the optical beams while retaining to the utmost the light emanating from faint sources in close proximity to the central star. At infrared wavelengths, the flux ratio between the star and the planet is more favourable than in the visible domain which makes nulling interferometers particularly attractive for the detection of light from exoplanets.
   \smallbreak
   We turn to a nulling architecture and data reduction technique which follows the basic concept devised by \cite{1978Natur.274..780B} and offers a reliable high-contrast imaging solution robust to small instrumental perturbations. This architecture, called kernel-nulling was proposed by \cite{2018A&A...619A..87M} and has been designed to exploit the use of a four-telescope interferometer. The two-stage kernel-nuller architecture produces three observable quantities which are accordingly referred to as kernel-nulls. Akin to kernel-phases \citep{2010ApJ...724..464M} which are an extension of closure phases \citep{1958MNRAS.118..276J}, kernel-nulls are independent of the quadratic error terms induced either by fluctuations in the amplitude or phase, but not a combination thereof. According to \cite{2004Lay}, the amplitude-phase cross terms are expected to be the dominant contributor to contrast detection limits. In combination with a photometric monitoring of the input beams, a kernel-nuller makes it possible to disentangle genuine weak astrophysical signals from spurious residual instrument-induced light leakage, enabling higher contrast detections in the presence of residual fringe tracking errors. Imperfections in the manufacturing and chromatic effects that would result in systematic errors can be compensated with a tunable recombiner that would follow the adaptive-nuller concept introduced by \cite{2003Lay}.
   \smallbreak
   In the paper that introduced the concept of kernel-nuller, \cite{2018A&A...619A..87M} computed transmission maps for a hypothetical snapshot scenario at zenith that did not attempt to account for the Earth rotation and its impact on the observable quantities. In this work, we take into account the Earth's rotation, describing how it affects such maps and actually improves the detectability of off-axis companions. We will also show how one can translate recorded series of kernel-nulls into constraints on the properties of a target. As the transmission maps are dependent on the layout of the telescopes and the pointing of the target, this study will lead us to comment on the characteristics of a possible observing program for the discovery of exoplanets with a kernel-nuller at the focus of the VLTI.

\section{The response of a nulling interferometer}
\label{sec:theory}

   \begin{figure*}
   \centering
   \includegraphics[scale=0.33]{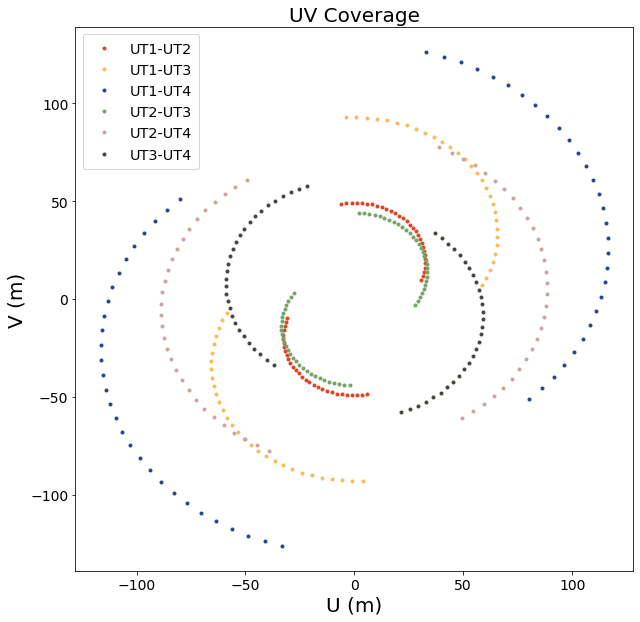}
   \includegraphics[scale=0.3]{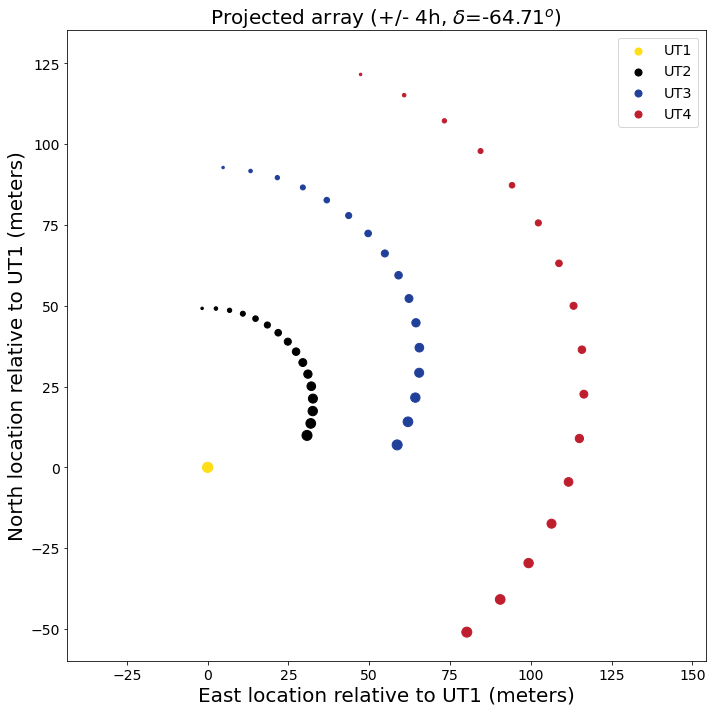}
   \caption{Evolution of the $(u, v)$ plane coverage (\emph{Left Panel}) and the projected array (\emph{Right Panel}) of the four VLTI UTs observing the target HIP 107773 over a $\pm 4$ hr range of hour angle. In the left figure we see the 2-D projection of the six baselines of the four UTs on a plane which is perpendicular to the direction of the target. As the Earth rotates, the baselines formed by pairs of telescopes sample the complex visibility function at spatial frequencies labelled with coordinates $u$ and $v$, sweeping out tracks which are rather circular on account of the favourable celestial location. In the right figure we see the projected geometry of the interferometric array relative to UT1 which is chosen as a reference. The locations of UT2, UT3 \& UT4, from the point of view of the source, appear to move around to UT1 in a clockwise direction.}
              \label{HIP 107773 - UV}%
    \end{figure*}

%
   \begin{figure}
   \centering
   \includegraphics[width=0.49\textwidth]{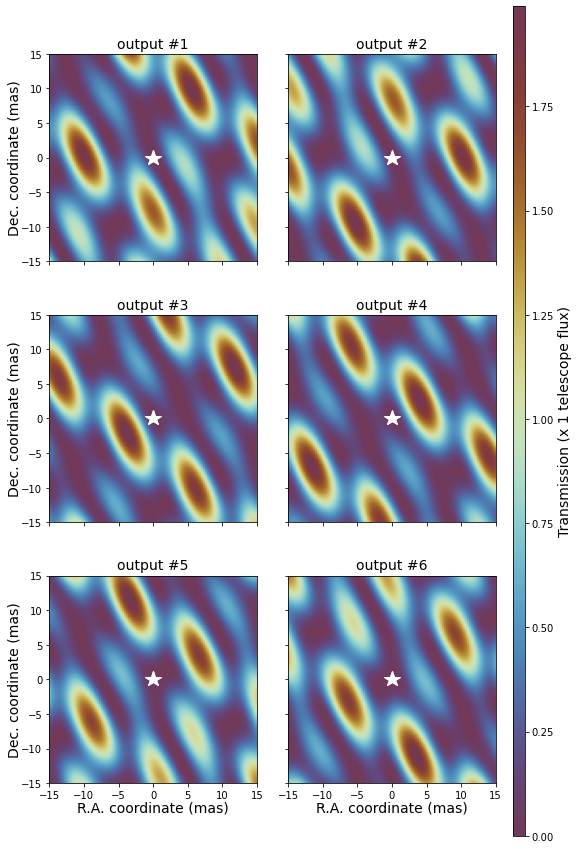}
      \caption{Nuller output transmission maps of a kernel-nuller at the focus of the four VLTI UTs, observing a $\pm$15 mas field of view surrounding the target HIP 107773. The figure shows the maps of the six nuller outputs for a pointing exactly at zenith. We see that these maps differ from one another and this diversity in the output response provides a constraint on the properties of a potential companion in the vicinity of a much brighter star. A white star in each map marks the location of the central star where the transmission for all channels, by design, is equal to zero. About the centre of the field, the maps are asymmetrical. For transmission maps, this asymmetry is a desirable trait as it allows for the elimination of centro-symmetric astrophysical features that would otherwise hide a planetary companion. With four input beams being split into six outputs, we expect an average planet flux of 4/6 (= 0.66) per channel. All six maps share the same colour scale where the transmission peaks at $\sim 2$ times the planet flux collected by a single aperture.}
         \label{sens_output}
   \end{figure}
   
   \begin{figure*}
   \centering
   \includegraphics[scale=0.3]{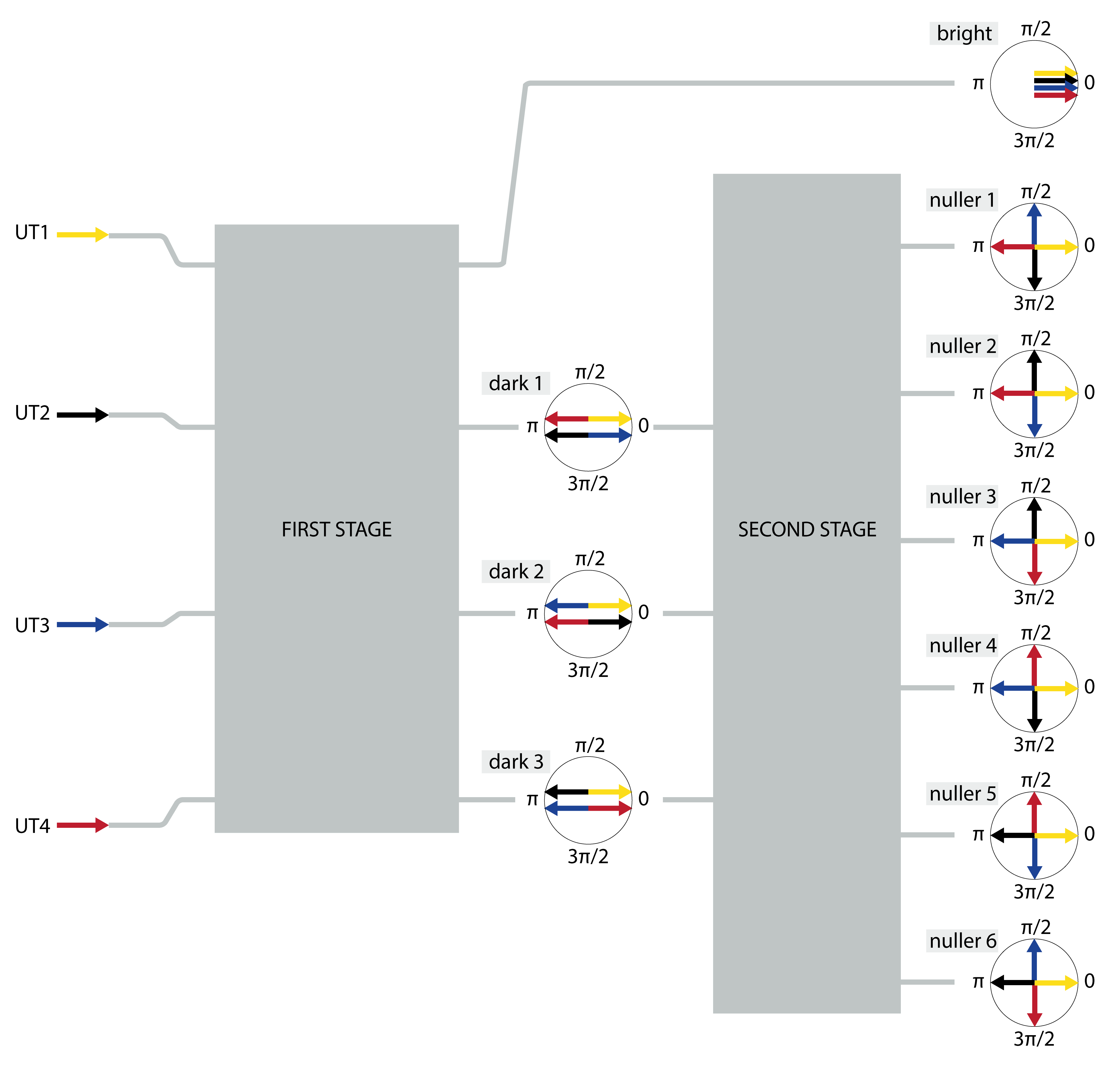}
   \caption{Schematic representation of the two-stage kernel-nuller architecture. Light from the four VLTI UTs is coupled into the first 4x4 stage that produces four outputs. Four in-phase ($\varphi = 0$) signals interfere constructively to produce a bright output (where most of the on-axis starlight is directed). For the three dark outputs, a phase shift ($\varphi = \pi$) is applied to the signal from two out of the four inputs. This means that when the light is combined, one input is out of phase from the other and so destructive interference occurs, leaving no starlight. At each stage of the four-beam combination, the different input (represented by coloured arrow) phase shifts applied to obtain the respective outputs are shown with complex matrix plots \citep{2020Laugier}. The second 3x6 stage performs six possible pairwise combinations of the dark outputs, with a phase shift ($\varphi = \frac{\pi}{2}$) introduced on one of the outputs. From the complex matrix plots of these six nuller outputs, we can see that pairs (nuller 1-2, 3-4, and 5-6) of outputs are mirror images of each other, an attribute that allows them to form a kernel-null.}
              \label{nuller_schematic}%
    \end{figure*}

   A conventional constructive interferometer recombines coherently the optical beams from two or more telescopes to measure the so-called complex visibility of an astronomical scene. Baselines formed by pairs of telescopes sample the complex visibility function at spatial frequencies labelled with coordinates $u$ and $v$. The accumulation of sufficient complex visibility measurements eventually leads to enough constraints that an image or a parametric representation of the astronomical scene can be reconstructed by virtue of the Van Cittert-Zernike theorem.
   \smallbreak
   Optical interferometers use a sparse array of telescopes which leads to an incomplete $(u, v)$ plane coverage in the snapshot mode. Hence observing facilities take advantage of aperture synthesis techniques which were first developed for radio astronomy in the late 1940s and have since been elaborated for the optical regime (see \cite{labeyrie_lipson_nisenson_2006} for an analysis on aperture synthesis for optical interferometry) to improve the $(u, v)$ plane coverage by making use of the Earth's rotation. As the Earth rotates, the projected baseline vector - as seen from the source - and the corresponding spatial frequencies vary with time, sweeping out tracks in the $(u,v)$ plane. Such tracks are conveniently summarised by a $(u,v)$ coverage plot (left panel of Fig.~\ref{HIP 107773 - UV}) whose overall density and distribution is a good diagnosis of the richness of the observation.   
   \smallbreak
   The information collected by a nuller is used in a different manner. In nulling interferometry, the requirement to redirect away the on-axis stellar light implies that the complex visibility of the source brightness distribution is in general not directly accessible. A four-beam nuller records the observed flux of a faint off-axis source acquired by a finite number of nulled outputs after the light from an on-axis source has been optically redirected toward one or more bright outputs. Even in ideal observing conditions, the link between the nulled outputs and the astronomical source is no longer described by a direct, linear relation such as the Van Cittert-Zernike theorem and the subsequent interpretation requires additional care. 
\subsection{Nuller output transmission maps}
   The most direct way to describe the non-trivial effect of a nuller recombining more than two telescopes is to compute transmission maps describing the response of the different outputs to the presence of a test point source over the field of view of the interferometer. These transmission maps  predict what fraction of the total flux of an off-axis source will find its way through the nuller, trickling down the different outputs. 
   \smallbreak
   One example of such maps is represented in Fig.~\ref{sens_output}, for the kernel-nuller architecture described by \cite{2018A&A...619A..87M}, and computed with the four unit telescopes (UTs) of the VLTI in the L-band ($\lambda = 3.6\: \mu m $) for a hypothetical scenario of a pointing exactly at zenith. To better illustrate the effects of nulling close to the central star, the field of view of a single snapshot in these maps has been set by the shortest ($b = 46.6$ meter) baseline of the array ($\lambda/b \simeq 15\: mas$). The recent experimental demonstration of a 3x3 kernel-nuller \citep{2022Cvetojevic} had a spectral resolution of 5.2 nm over a 0.55 $\mu m$ bandwidth, which we adopt for this investigation.  
   \smallbreak
   At a glance, we can see that these maps are not uniform and differ from one another. Other than the common on-axis transmission hole, which is the very purpose of the nuller, each map features bumps and gaps at distinct locations which are imposed both by the arrangement of the telescopes on the ground and by the inner details of the recombiner itself. Differences in these maps are a desirable feature: the measurement brought by each output is a new piece of information that will constrain the spatial distribution of intensity of the target astrophysical scene and one noteworthy feature of this particular architecture of nuller is that it produces as many distinct outputs as the total number of baselines of the interferometer. The sparse nature of the interferometer however also makes that the monochromatic nuller output transmission maps are periodic. A blind snapshot observation with a nuller may very well by chance lead to the direct detection of the flux of an off-axis source, but will not suffice to unambiguously locate the source in the field of view. In this work, just like what is done with aperture synthesis of constructive interferometry, we will take advantage of the diversity brought by the rotation of the Earth and its effect on the different outputs, and their time-evolving and pointing dependent transmission maps.   
   \subsection{Projected interferometric array}
   Instead of having to keep track of the evolving $(u, v)$ coordinates of the six baselines in the plane perpendicular to the line of the interferometric pointing (see the left panel of Fig.~\ref{HIP 107773 - UV}), the computation of the nulled output transmission maps requires only to keep track of the projected geometry of the interferometric array, relative to one sub-aperture arbitrarily chosen as a reference. We can use the same coordinate transformation matrix (see for instance \cite{2007NewAR..51..597S}) that computes the instantaneous $(u, v)$ coordinates of an interferometer located at the mean latitude $l$ and pointing at declination $\delta$ and hour angle $h$: 
   
    \begin{gather} 
    \label{proj_uv_track}
 \begin{pmatrix} u \\ v \end{pmatrix}
 =
  \begin{pmatrix}
   -\sin(l) \sin(h) &
   \cos(h) \\
   \sin(l) \cos(h) \sin(\delta)+\cos(l) \cos(\delta) &
   \sin(h) \sin(\delta) 
   \end{pmatrix}
 \begin{pmatrix} B_{\mathrm{north}} \\ B_{\mathrm{east}} \end{pmatrix}, 
\end{gather}

\noindent but only to compute the apparent location of the three sub-apertures of the VLTI that, from the point of view of the source, appear to move around to the reference sub-aperture. The right panel of Fig.~\ref{HIP 107773 - UV} shows how this projected interferometric array evolves over time, with UT1 chosen as a reference, and compares it to the usual $(u, v)$ coverage plot visible to the left. These telescope coordinate tracks were computed over an hour angle range of $\pm 4$ hr for the specific example of HIP 107773, located at a favourable celestial location ($\delta= -64 ^\circ 42' 45''$) for the VLTI ($l= -24 ^\circ 24' 37''$) which results in rather circular tracks. We will keep using this example to see the effect this has on the outputs of a kernel-nuller at the focus of the VLTI. 
   
 \subsection{Time evolving output transmission maps}
   At any instant, using the matrix transform of Eq. \eqref{proj_uv_track}, we can compute the apparent $(x, y)$ coordinates of the UTs relative to UT1. To compute the output transmission maps across the field of view, we need to keep track of the value of the electric field emitted by an off-axis test source of right ascension and declination offset ($\alpha$, $\delta$) and collected by the different sub-apertures, possibly affected by some amount of residual piston $\rho$. For this point source, the 4-element vector $U$ (commensurate with the number of UTs) of complex amplitudes at the aperture entrance with wavelength $\lambda$ is: 
   
   \begin{equation}U(\alpha,\delta) \propto e^{(-2\pi i/\lambda) \cdot (\rho+(x\alpha+y\delta))} \label{u_vect} \end{equation}
   
   \noindent Having taken UT1 as the spatial reference, it is natural to also select it as the phase reference so that the phases of the different electric fields sampled by the other UTs are measured relative to this reference. The piston values are also quoted relative to UT1. The coherent interferometric combinations performed between the sub-apertures are fully described by a complex matrix \textbf{M} which links the input complex amplitudes of the off-axis test source to the complex amplitudes of the nulled output. A detector records the intensity (i.e., square modulus) of the complex amplitude. We will refer to the two-dimensional response of the nuller outputs at time $t$:
   
      \begin{equation}   \tau(\alpha, \delta, t) =\: \mid \textbf{M}\cdot U(\alpha, \delta, t) \mid^2 \label{omap}  \end{equation}

  \noindent as nuller output transmission maps, and we will use them extensively to comment on the properties of several observing scenarios with the kernel-nuller. At any instant, for an astrophysical scene described by a (usually unknown) function $O(\alpha, \delta)$, the raw output effectively recorded by the nuller will be the six-component vector: 
    
   \begin{equation} 
   I(t) = \int \tau(\alpha, \delta, t)\cdot O(\alpha, \delta) \,\mathrm{d}\alpha \mathrm{d}\delta \label{intsy} \end{equation}
   
   Kernel outputs, noted $\kappa$, which are the primary intended observable quantities for the kernel-nuller, are linear combinations of simultaneous raw outputs. These combinations are summarised by a left hand operator $\mathbf{K}$, that results in the 3-element vector: $\kappa(t) = \mathbf{K} \cdot I(t)$ . In Sec. \ref{sec:colin}, we will see how to go from a record of raw outputs $I(t)$ or kernel outputs $\kappa(t)$ back to a description of the object $O(\alpha, \delta)$. To comment on some specific properties of the kernel-nuller, it will also be useful to look at how the kernel outputs respond to the presence of an off-axis point source. We therefore also introduce kernel output maps $\kappa_M(\alpha, \delta, t)$ that will be directly computed as: 
   
   \begin{equation} \kappa_M(\alpha, \delta, t) = \textbf{K}\cdot \tau(\alpha, \delta, t) \label{kmap} \end{equation}
   
\section{Simulated kernel-nulling observations at the VLTI}
%
   \begin{figure*}
   \centering
   \includegraphics[width=\textwidth]{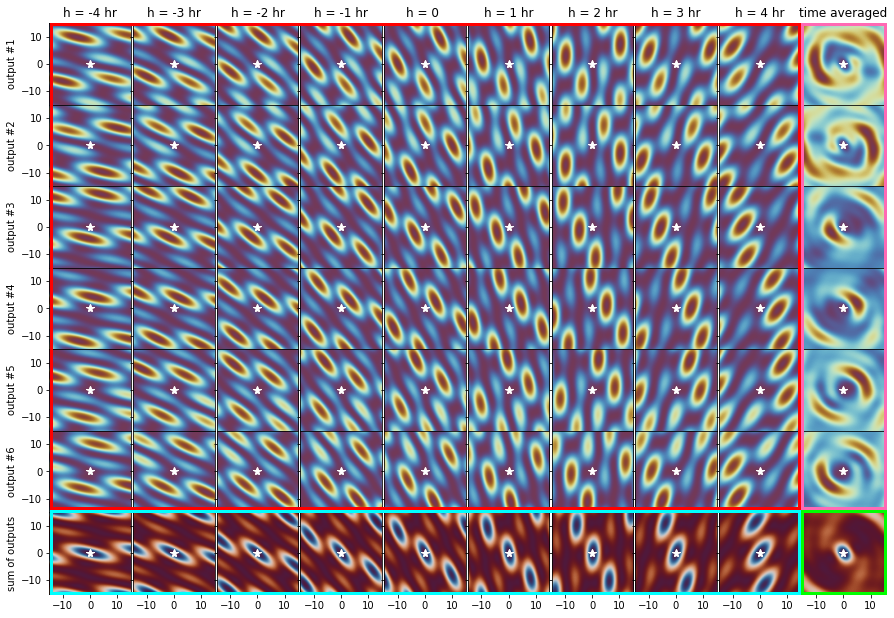}
   \caption{Evolution of the nuller output transmission maps of a kernel-nuller at the focus of the four VLTI UTs, observing the $\pm$15 mas field of view surrounding the target HIP 107773. The figure shows all together four blocks of maps that provide insights into the behaviour of the nulled outputs. The first block (red contour, top left) of 54 maps shows how the transmission maps of the six nuller outputs evolve over a $\pm 4$ hour observing period centred on the target transit time ($h=0$). These maps all share the same colour scale as Fig.~\ref{sens_output} and show that at any instant, the transmission for at least part of the field of view per output can reach up to the totality of the flux collected by two telescopes. We can see the transmission patterns gradually rotating in a clockwise direction with time, scanning the field of view to intercept the light of any off-axis feature of the astrophysical scene. A white star in each plot marks the location of the central star where the transmission for all channels, by design, is equal to zero. The second block (pink contour, right most column) of 6 maps shows, using the same colour scales and conventions, the time-average of each nuller output transmission map. Looking at these, we can visually confirm that the outputs do indeed work in pairs characterised by asymmetric responses. The third block (blue contour, bottom row) of 9 maps shows how at any instant, the sum of all output maps is spatially distributed. This series of images uses the same colour scale as Fig.~\ref{NGTM+CMAP}. Adding all together the outputs shows that the overall instantaneous sensitivity of the nuller is much more uniform over the field of view, even if it nevertheless systematically features off-axis replicas of the on-axis null, which are referred to as transmission holes. The fourth block (green contour, single bottom right map) shows the time-averaged global throughput. It shows that with a sufficiently wide observing window, the field of view is uniformly covered, and the nuller will be able to capture the light of a companion regardless of where it may hide.}
              \label{Nmap_evo}%
    \end{figure*}
   \begin{figure*}
   \centering
   \includegraphics[width=\textwidth]{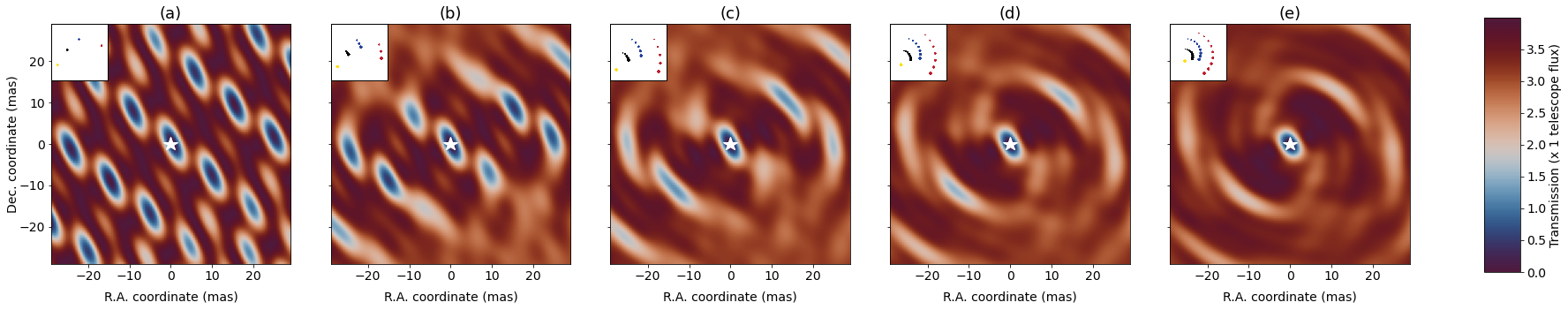}
   \caption{Nuller global throughput maps at various observation periods of a kernel-nuller at the focus of the four VLTI UTs, observing the $\pm$30 mas field of view surrounding the target HIP 107773. The throughput maps allow us to assess the overall off-axis efficiency of a nulling observing sequence. In Map (a) we see the throughput map for a single snapshot pointing at zenith. For the following maps, we computed the throughput maps at each pointing and then co-added the maps over the range of hour angle: in Map (b) we see the throughput map for three equally spaced pointings over a $\pm 1$ hr range of hour angle, Map (c) for five equally spaced pointings over $\pm 2$ hr, Map (d) for seven equally spaced pointings over $\pm 3$ hr and Map (e) for nine equally spaced pointings over $\pm 4$ hr. The pointings of the projected array are shown inset in each throughput map. A white star in each map marks the location of the central star where the rejection by the nuller is optimal as shown by the on-axis null. The five maps share the same colour bar where the transmission is expressed in units of the planet flux.}
              \label{NGTM+CMAP}%
    \end{figure*}

   \begin{figure*}
   \centering
   \includegraphics[width=\textwidth]{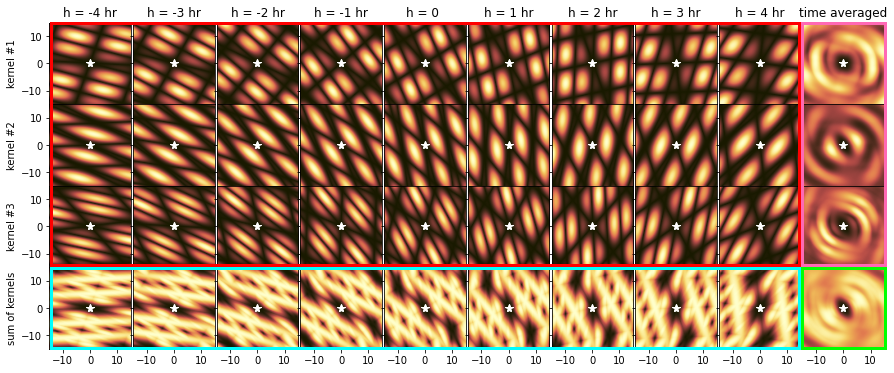}
   \caption{Evolution of the kernel magnitude maps of a kernel-nuller at the focus of the four VLTI UTs, observing the $\pm$15 mas field of view surrounding the target HIP 107773. The figure shows all together four blocks of maps that provide insights into the behaviour of the kernel-outputs. The first block (red contour, top left) of 54 maps shows how the magnitude maps of the three kernel-nuller outputs evolve over a $\pm 4$ hour observing period centred on the target transit time ($h=0$). We can see the transmission patterns gradually rotating in a clockwise direction with time, scanning the field of view to intercept the light of any off-axis feature of the astrophysical scene. A white star in each plot marks the location of the central star where the transmission for all channels, by design, is equal to zero. The second block (pink contour, right most column) of 3 maps shows the time-average of each kernel output magnitude map. The third block (blue contour, bottom row) of 9 maps shows how at any instant, the sum of all kernel magnitude maps is spatially distributed. Adding all together the kernel-outputs shows that the overall instantaneous sensitivity of the nuller is much more uniform over the field of view, even if it nevertheless systematically features off-axis replicas of the on-axis null, which are referred to as transmission holes. The fourth block (green contour, single bottom right map) shows the time-averaged global throughput. It shows that with a sufficiently wide observing window, the field of view is uniformly covered, and the nuller will be able to capture the light of a companion regardless of where it may hide. All maps share the same colour scale where the transmission is expressed in units of the kernel amplitude collected by one telescope.}
              \label{Kmags_kgtm}%
    \end{figure*}

    \begin{figure*}
   \centering
   \includegraphics[width=0.49\textwidth]{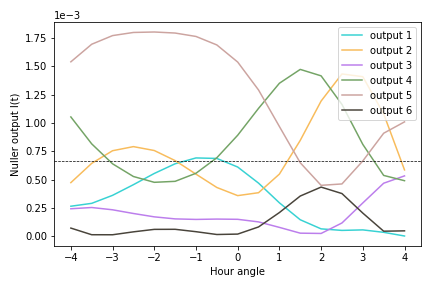}
   \includegraphics[width=0.49\textwidth]{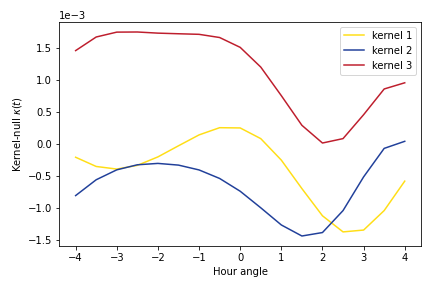}
   \caption{Modulation of the raw null (\emph{Left Panel}) and the kernel-null (\emph{Right Panel}) induced by a single off-axis companion located at $\alpha$ = +5.0 mas, $\delta$ = +5.0 and with a contrast $c = 10^{-3}$. This simulated kernel-nulling observation of HIP 107773 at the focus of the four VLTI UTs was in the absence of perturbations. The figures show that there is no redundancy in the kernel-nuller architecture as each of the six raw outputs and each of the three kernel outputs produces a different signal over the $\pm 4$ hr range of hour angle. Whereas the raw outputs are all positive, the kernel outputs can take negative values. If the action of the nuller was uniform, each channel would register a constant and uniform amount of flux, which in the units of the \emph{Left Panel} would be $2/3 \times 10^{-3}$ (as shown by the dashed line).}
              \label{response}%
    \end{figure*}

    In this section, we present simulated monochromatic kernel-nulling L-band ($\lambda = 3.6\: \mu m $) observations of the aforementioned target HIP 107773 at the focus of the four VLTI UTs. The simulated observing sequence consists of nine equally spaced pointings over the $\pm 4$ hr range of hour angle. We are going to look at the output and the properties of the baseline kernel-nuller design of \cite{2018A&A...619A..87M}.   
    \smallbreak  
    Fig.~\ref{nuller_schematic} is a schematic representation of this two-stage nuller architecture, which shows the 4-UT beam combination that results in six nuller outputs, with the different phase shifts applied at each stage. In the first 4x4 stage, four in-phase signals from the UTs are combined to produce a bright output (bright in Fig.~\ref{nuller_schematic}). For the three distinct dark outputs (dark 1-3 in Fig.~\ref{nuller_schematic}), pairs of UTs along each nulling baseline are combined with a $\pi$ phase shift on one of the input signals. To produce dark output 1, UT1 (set as the phase reference) is combined with UT4 and UT2 is combined with UT3. The combination of UT1-UT3 and UT2-UT4 yields dark output 2 whereas UT1-UT2 and UT3-UT4 generates dark output 3. The second 3x6 stage leaves the bright output untouched and performs six possible pairwise combinations of the dark outputs. By introducing a $\frac{\pi}{2}$ phase shift on one of the outputs, each dark output is split into two asymmetric responses, which is analogous to phase chopping \citep{1997ApJ...475..373A,2004Lay,2005Mennesson}. The overall effect of the recombiner taking four UTs and producing six nuller outputs (nuller 1-6 in Fig.~\ref{nuller_schematic}) is described by a 6x4 complex matrix \textbf{M}, that we use to compute the response of these outputs (see Eq. \eqref{omap}): 
    
    \begin{gather} 
 \textbf{M}
 = \frac{1}{4} \times
  \begin{bmatrix}
   1+j &
   1-j & -1+j & -1-j \\
   1+j &
   -1+j & 1-j & -1-j \\
   1+j &
   1-j & -1-j & -1+j \\
   1+j &
   -1+j & -1-j & 1-j \\
   1+j &
   -1-j & 1-j & -1+j \\
   1+j &
   -1-j & -1+j & 1-j \\
   \end{bmatrix}
\end{gather}

    \noindent To this nuller is also associated a kernel matrix $\mathbf{K}$ which, when applied to the raw outputs of the nuller, produces observable quantities called kernel outputs, intrinsically more robust to fringe tracking residuals:
   
    \begin{gather} 
 \textbf{K}
 = 
  \begin{bmatrix}
   1 &
   -1 & 0 & 0 & 0 & 0 \\
   0 &
   0 & 1 & -1 & 0 & 0 \\
   0 &
   0 & 0 & 0 & 1 & -1 \\
   \end{bmatrix} 
\end{gather}

    \noindent For this design, a kernel output is therefore simply the difference between the two nuller outputs for which the same pairs of telescopes are nulled with each other (nuller 1-2, 3-4, and 5-6 in Fig.~\ref{nuller_schematic}). Due to the asymmetric response of the nuller outputs (e.g., nuller 1-2 in Fig.~\ref{nuller_schematic}), a centro-symmetric signal is eliminated by the subtraction. A difference of output values also means that unlike raw photometric-like outputs, kernel outputs can take negative values. To further comment on the properties of the kernel-nuller, it will prove useful to examine the absolute value of the kernel output. In addition to monitoring the evolution of the nuller output transmission maps $\tau(\alpha, \delta, t)$ in Sec. \ref{sec:nmaps} and the global nuller throughput in Sec. \ref{sec:glomap}, we will therefore also look at kernel magnitude maps $\mid \kappa_M(\alpha, \delta, t) \mid$ (in Sec. \ref{sec:kmag}).
    
\subsection{Nuller output transmission maps with time}  
\label{sec:nmaps}

    The optimal configuration for a nuller is when the bright target is positioned at the centre of the transmission map and the off-axis feature is situated in an area that is mostly driven toward the nulled outputs. As the location of the off-axis feature may be unknown, its detection in a snapshot is not guaranteed. Even in case some signal is detected, we cannot directly tell where the flux is coming from, since the signal is the result of an integral over the field of view (see Eq. \eqref{intsy}). This is why \cite{1978Natur.274..780B} proposed to rotate the interferometer: the on-axis bright star remains nulled, while off-axis features travel through the transmission maps, resulting in intensity modulations on the different outputs that can later be interpreted.  
    \smallbreak
    With a ground based long-baseline interferometer, the scenario is different. For any pointing, in addition to the expected rotation of the transmission pattern manifest in Fig.~\ref{Nmap_evo}, the different baselines stretch and shrink over the course of the transit. This is made evident by the transmission patterns which have a broad appearance when the projected baselines are short (column 1 in the first block of Fig.~\ref{Nmap_evo}) and appear narrow when the baselines are long at transit (column 5 in the first block of Fig.~\ref{Nmap_evo}). As the source moves away from transit, the baselines begin to shorten again and we notice a progressive increase in the size of the transmission patterns. 
    \smallbreak
    From the time-averaged transmission maps shown in the second block of Fig.~\ref{Nmap_evo} we can confirm that the raw outputs do indeed work in pairs. If we isolate the time-averaged outputs 1 \& 2 we see that the resulting dark spiral tracks in output 2 occupy the previous locations of bright spiral tracks in output 1 and vice versa. The same applies for outputs 3 \& 4, and outputs 5 \& 6. Consequently, a potential off-axis source located on the maximum of transmission of one map would not give any signal through the other map. Thus if we alternately detect a pair of outputs, the signal of an off-axis source can be further modulated and therefore retrieved.
    \smallbreak
    The optimal sensitivity for each of these pairs of outputs occurs at slightly different angular separations in right ascension offset: $\pm 10$ mas for outputs 1 \& 2, $\pm 3$ mas for outputs 3 \& 4 and $\pm 6$ mas for outputs 5 \& 6. This is in line with the length of the shortest nulling baselines used for each pair of outputs, the shortest one being used for outputs 1 \& 2 (UT2-UT3 = 46.6-m) and the longest one for outputs 3 \& 4 (UT2-UT4 = 89.4-m). On average, over the entire $\pm 4$ hour observing window, these lead to a rather uniform sensitivity. At any instant, the individual nuller outputs have an average throughput of $\sim 1/6$th of the total flux collected by four telescopes. 
    
 \subsection{Nuller global throughput map with time}
 \label{sec:glomap}
 
    To assess the efficiency of the nuller at any given time for a specific observing scenario, it is useful to monitor the evolution of the overall throughput. Adding together the six nuller output transmission maps at any instant yields a new map that we refer to as the nuller global throughput map (third block of Fig.~\ref{Nmap_evo}). In the absence of prior information (ephemeris) about the location of a potential off-axis companion, the throughput map (time averaged over the planned observing window) can be used to predict the probability of detection. Located at the centre of these maps is an on-axis null which upon observation is seen to go through a $\sim90^\circ$ clockwise rotation over the $\pm 4$ hr range of hour angle . Replicas of the on-axis null (transmission holes) are visible in the throughput map. The layout of the four VLTI UTs generally favours detections in the top-right and bottom-left corners of the maps, however, Earth rotation eventually results in a more uniform coverage of the throughput map as shown in the fourth (bottom right) block of Fig.~\ref{Nmap_evo}.
    \smallbreak
    The extent of the maps we have looked at thus far has been limited to the field of view provided by the shortest snapshot baseline. The effective field of view (FOV) extends to the diffraction limit of a UT ($\lambda/d \simeq 90$ mas): any structure located within this FOV will be coupled into the recombiner and it is therefore relevant to look at what is happening for an extended FOV. 
    \smallbreak
    Panel (a) of Fig.~\ref{NGTM+CMAP} reproduces the throughput map for a snapshot observation at zenith, over an extended $\pm30$ mas FOV. The angular separation at which the on-axis null first reaches half of the peak throughput can be used as an approximation of the inner working angle (IWA). For the scenario at zenith, the peak throughput recorded is $\sim3.9$, which gives an IWA of $\pm 4$ mas in right ascension offset and $\pm 6$ mas in declination offset. The periodic response of the nuller in this monochromatic scenario is made obvious with the numerous replicas of the on-axis null - that would hide the signal of a companion - repeating every 15 mas. As the size of the observing window progressively increases, the transmission holes eventually fill in, with the nuller exhibiting a more homogeneous throughput after an eight hour observation period (Panel (e) of Fig.~\ref{NGTM+CMAP}). The peak throughput at this range of hour angle is $\sim3.7$ and the IWA is observed to be $\pm 4$ mas in right ascension and declination offset. Earth rotation does change things somewhat for a suitable target like HIP 107773, but less favourable scenarios at higher declination angles will retain some transmission holes in the throughput maps even with a sufficiently wide observing window.
    
   \subsection{Kernel magnitude maps with time}
    \label{sec:kmag}
    
    The primary motivation for the kernel-nuller is the production of observable quantities called kernel-nulls that are less sensitive to residual piston fluctuations. We've seen that for this specific architecture, kernel outputs are simply the pairwise difference between consecutive raw outputs (1-2, 3-4, and 5-6). Whereas raw outputs are a measure of flux, which are always positive, kernel outputs can take negative values (see Fig.~\ref{response}). So in order to make quantitative observations about the detectability by a kernel-nuller, we will look at kernel magnitude maps.
    \smallbreak
    The kernel magnitude maps presented in the first block of Fig.~\ref{Kmags_kgtm} are analogous to the nuller output transmission maps, however, some differences between the respective maps are apparent. By virtue of taking the norm of the kernel output we observe a symmetry about the centre of the field in the magnitude maps and we also detect an increased number of transmission patterns. This means that for a given observation period, a potential companion will travel through more successive bright and dark regions in the magnitude maps thus generating higher harmonics in the detected signal which can give a stronger constraint on the position of the companion. In Sec. \ref{sec:colin}, we will show how this signal can be used to give a description of a companion around a bright star. 
    \smallbreak
    From the time-averaged magnitude maps shown in the second block of Fig.~\ref{Kmags_kgtm} we observe that the kernel outputs display a more homogeneous sensitivity in the field of view relative to the raw outputs. Each output reaches its optimal sensitivity at varying positions in right ascension offset: with output 1 at $\pm 10$ mas, output 2 at $\pm 3$ mas and output 3 at $\pm 6$ mas. In the third block of Fig.~\ref{Kmags_kgtm} we have computed throughput maps which are referred to as kernel global throughput maps and we observe from the time-averaged throughput map shown in the fourth block of Fig.~\ref{Kmags_kgtm} that with a sufficiently wide observing window, the field of view is uniformly covered.

   \section{Translation of kernel-nuller signals}
      
   \begin{figure*}
   \centering
   \includegraphics[width=1.0\textwidth]{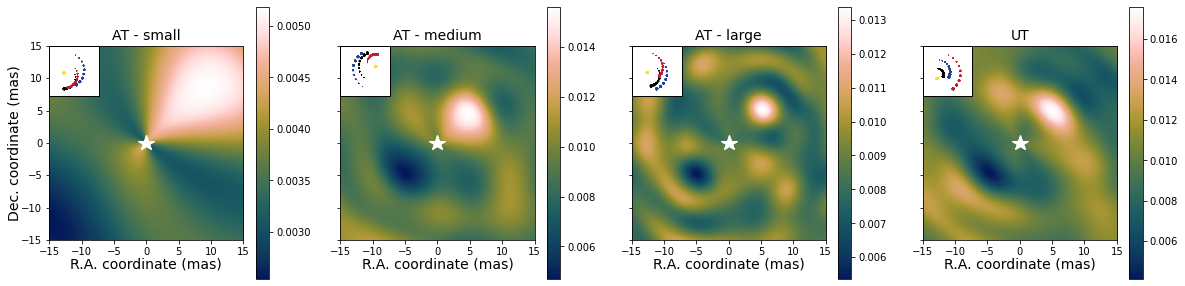}
   \caption{Colinearity maps for different four telescope VLTI interferometric array configurations over a $\pm$15 mas field of view. These maps have been computed from the data-set acquired from simulated kernel-nulling observations of HIP 107773 over a $\pm 4$ hr range of hour angle. For these simulations in the absence of perturbations, an off-axis companion with a contrast $c = 10^{-3}$ was placed at $\alpha$ = +5.0 mas, $\delta$ = +5.0. For the \emph{small} (A0-B2-D0-C1) configuration of the ATs we observe a maximum of colinearity confined to the top right of the map. Owing to the low spatial resolution provided by this array, we cannot infer the exact location of the companion. The colinearity map is improved with the use of the \emph{medium} (K0-G2-D0-J3) AT configuration. Here, we see a broad maximum of colinearity around the location of the companion. The longer baselines provided by the \emph{large} (A0-G1-J2-J3) AT configuration and the \emph{UTs} will result in a higher spatial resolution. This is made evident in the colinearity maps computed with these array configurations by a narrower maximum of colinearity around the location of the companion. The scale on the right of each colinearity map denotes the amplitude of the companion signal which is proportional to its contrast. The pointings of the projected array are shown inset in each colinearity map. A white star in each map marks the location of the central star where the rejection by the nuller is optimal.}
    \label{colinearity_map}%
    \end{figure*}

    \begin{figure*}
   \centering
   \includegraphics[scale=0.32]{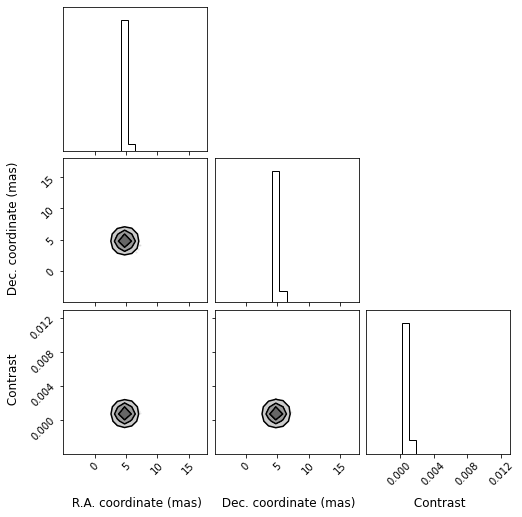}
   \includegraphics[scale=0.32]{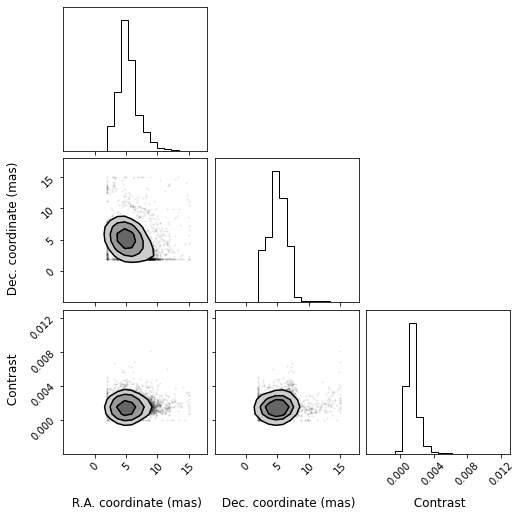}
   \includegraphics[scale=0.32]{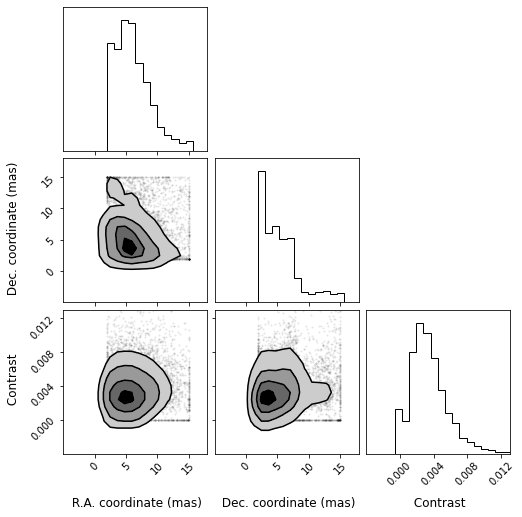}
   \caption{Corner plots of the distribution of fitted values in position ($\alpha$, $\delta$) and contrast $c$ for an off-axis companion located at $\alpha = +5.0$ mas, $\delta = +5.0$ mas and with a contrast $c = 10^{-3}$. The figures show the distributions of the respective parameters at various amounts of piston residuals. \emph{Left Panel:\/} At 50 nm we see well distributed values around the coordinate locations of the companion with $\sigma_{\mathrm{RA}} = 0.22$ mas \& $\sigma_{\mathrm{DEC}} = 0.25$ mas; \emph{Middle Panel:\/} At 100 nm we begin to see some flux leakage from the null particularly in the $\delta$ distribution with $\sigma_{\mathrm{RA}} = 2.01$ mas \& $\sigma_{\mathrm{DEC}} = 2.26$ mas; \emph{Right Panel:\/} At 150 nm, the piston-induced leaked intensity is more pronounced for both $\alpha$ \& $\delta$ with $\sigma_{\mathrm{RA}} = 4.11$ mas \& $\sigma_{\mathrm{DEC}} = 4.30$ mas. Each 4-UT MC simulated data-set consists of $10^4$ series of kernel outputs acquired in the presence of a controlled amount of residual phase noise.}
              \label{Least squares}%
    \end{figure*}
    
    In this section we will see how the signals acquired by a kernel-nuller can be exploited to build up a useful two-dimensional representation of a target of interest. Section \ref{sec:theory} introduced the formalism required to account for the effect of Earth rotation on the raw outputs $I(t)$ produced by the nuller and their subsequent kernel outputs $\kappa(t)$. Eq. \eqref{intsy} in particular describes how, in the presence of Earth rotation, the object function $O(\alpha, \delta)$ describing the target becomes encoded in temporal fluctuations of the recorded raw outputs $I(t)$. To illustrate this effect, we performed a simulated observing sequence of HIP 107773 consisting of seventeen equally spaced pointings over a $\pm 4$ hr range of hour angle. The simulation includes an off-axis companion conveniently located at $\alpha = +5.0$ mas, $\delta = +5.0$ mas (for a contrast $c = 10^{-3}$), where (according to Fig.~\ref{Nmap_evo} and Fig.~\ref{Kmags_kgtm}) the sensitivity of the 4-UT configuration of the VLTI is expected to be optimal. The presence of the companion induces a modulation of the different outputs of the nuller and their subsequent kernels, respectively represented in left and right panels of Fig.~\ref{response}. If the action of the nuller was uniform, each channel would register a constant and uniform amount of flux, which in the units of the left panel of Fig.~\ref{response} would be $2/3 \times 10^{-3}$. In this favourable location, some outputs can record up to twice that amount.

\subsection{Colinearity map}
\label{sec:colin}
     
     To decode these temporal fluctuations of the nuller measurements and turn them back into constraints on the spatial structure of the object, we can correlate the sequence of signals $I(t)$ acquired during an observation with the response of the nuller $\tau(\alpha, \delta, t)$ computed for the target over the same observation window, which produces a 2D map:
    
     \begin{equation} C_{\mathrm{I}}(\alpha, \delta) =\frac{\int I(t) \cdot \tau(\alpha, \delta, t) \,\mathrm{d}t.}{\int \tau(\alpha, \delta, t) \,\mathrm{d}t.} \label{colinN} \end{equation}
    
     \noindent This procedure is identical to the concept of a 2D binary model colinearity map used by \cite{2019A&A...623A.164L} \& \cite{2020A&A...636A..72M} in the context of kernel-phase interferometry. The recorded signals $I(t)$ or their kernel outputs $\kappa(t)$ are effectively projected onto a 2D grid of possible binary models. Assuming that we get enough diversity over time in the signal and a nuller transmission response that is unambiguous - as is the case with a kernel-nuller producing asymmetric outputs - the resulting map can serve as a first estimate for the object function $O(\alpha, \delta)$ (see Sec.~\ref{sec:theory}) from which the bright on-axis source has been removed. Colinearity maps can be constructed from the raw nuller outputs (Eq. \eqref{colinN}) as well as from the kernel of these outputs: 
     
     \begin{equation} C_{\mathrm{\kappa}}(\alpha, \delta)  =\frac{\int \kappa(t) \cdot \kappa_M(\alpha, \delta, t) \,\mathrm{d}t.}{\int \kappa_M(\alpha, \delta, t) \,\mathrm{d}t.}\label{colinK} \end{equation}

     \noindent as described in Sect. 3 which now include an off-axis companion (presented in Fig.~\ref{response}) located at $\alpha = +5.0$ mas, $\delta = +5.0$ mas and with a contrast $c = 10^{-3}$. One can verify that the sensitivity of the nuller with the raw outputs and kernel outputs is optimal at these coordinates with the throughput maps presented in the bottom right of Fig.~\ref{Nmap_evo} and Fig.~\ref{Kmags_kgtm}. 
     \smallbreak
     The fidelity of the colinearity map is dependent on the spatial resolution provided by the interferometric array. The larger the baselines of the array, the higher the spatial resolution, thereby improving the accuracy on the estimate of $O(\alpha, \delta)$. For observations with four telescopes, the VLTI can operate with the 8-m UTs or with the 1.8-m auxiliary telescopes (ATs). Whilst the positions of the UTs are fixed, the ATs are relocatable and can be moved to a number of different stations. This way, observations on several different baselines can be carried out. The list of AT quadruplets offered in P110 (1 October 2022 – 31 March 2023) are:
     
     \begin{itemize}
      \item small, consisting of stations A0-B2-D0-C1, baseline lengths between 11.3 to 33.9-m.
      
      \item medium, consisting of stations K0-G2-D0-J3, baseline lengths between 39.99 to 104.3-m.
      
      \item large, consisting of stations A0-G1-J2-J3, baseline lengths between 58.2 and 132.4-m.
   \end{itemize}
     
    \noindent  Fig.~\ref{colinearity_map} shows the result of the computation of several colinearity maps $C_{\mathrm{I}}(\alpha, \delta)$ over a grid of possible locations covering a $\pm15$ mas FOV for the 3 AT configurations and the UTs. As expected, the shorter baselines of the small configuration have a low spatial resolution therefore we see in the colinearity map of this array that $O(\alpha, \delta)$ is estimated over a wide region encompassing $\sim15$ mas in right ascension and declination offset. The large configuration offers the longest baselines and in the map of this array we observe a narrower spatial structure of $O(\alpha, \delta)$ meaning that we can infer more accurately the location of the companion. The colinearity maps derived with the medium, large and UT configurations suggest that the use of these arrays can allow us to effectively demodulate the signal of the companion. In these maps, a maximum of colinearity can be observed where the companion is expected, and since the amount of flux present in the different outputs is proportional to the brightness of the companion, when properly normalised, the maximum of colinearity matches the contrast of the companion. 
     
\subsection{Extracting astrometric parameters of an off-axis companion}
\label{sec:astrometry}
    
    The location of the maximum of colinearity in the map shown in Sec.~\ref{sec:colin} of the 4-UT configuration can be used as a first estimate for the position ($\alpha$, $\delta$) and contrast $c$ of the actual companion. This first estimate can then serve as the starting point in a parametric $\chi^2$ minimisation procedure where the data (real or simulated) is compared to the parametric model of a binary object. The introduction of piston phase affects the fidelity of the fit. To estimate how these residual fringe tracker errors affect our relative astrometry, we use Monte Carlo (MC) simulations. Each MC simulated data-set consists of $10^4$ series of kernel outputs acquired in the presence of a controlled amount of residual phase noise. We use the $lmfit$ python package \citep{2016ascl.soft06014N} which is an implementation of the Levenberg-Marquardt algorithm to find the best fitting binary parameters and look at how they're distributed. Here, $\chi^2$ is a measure of the mean-squared difference between the measured kernel values $\kappa(t)$ and the corresponding values for the model $\kappa_B(\alpha, \delta, c, t)$:   
    
     \begin{equation} \chi^2(\alpha, \delta, c) =  \sum_t \ \frac{\mid \kappa(t) - \kappa_{\mathrm{B}}(\alpha, \delta, c, t) \mid^2}{\sigma^2}, \end{equation}

    \noindent where $\sigma^2$ is the variance of the piston phase in $\kappa(t)$. Fig.~\ref{Least squares} shows the result of some of these MC simulations for 50, 100 and 150 nm of fringe-tracking residuals. To put these values into context, current levels of VLTI/GRAVITY fringe tracker residuals are $\sim200$ nm on the UTs though the GRAVITY+ upgrade will aim to go below 100 nm RMS. When deployed on telescopes, a nuller is extremely sensitive to such phase perturbations even after adaptive optics or fringe tracker correction. The self-calibrating observables derived from a kernel-nuller attempt to alleviate this and we observe from the right panel of Fig.~\ref{Least squares} that even with 150 nm piston induced error, the variance in the respective astrometric parameters is still relatively low.

\section{Kernel-nuller exoplanet discovery observing program}

\begin{table*}
	\begin{center}
		\begin{tabular}{c c c c c c c c c c c c c}
		\hline \hline
			HIP & R & K & SpT & $\varpi$ & $\Delta v_\mathrm{T,G3}$ & $m_{\mathrm{\star}}$ & $R_{\mathrm{\star}}$ & $m_{\mathrm{P}}$ & Sep. & $\log(c)$ & Assoc. & Age \\
			 & (mag) & (mag) & & (mas) & (m/s) & ($M_{\mathrm{\odot}}$) & ($R_{\mathrm{\odot}}$) & ($M_{\mathrm{Jup}}$) & (mas) & & & (Myr) \\
			\hline
			1481 & 7.13 & 6.15 & F8V & 23.35 & 22.13 & 1.16 & 1.12 & 2.55 & 116.78 & -6.78 & THA & 45 \\
			14157 & 7.92 & 6.55 & K0V & 19.45 & 69.06 & 0.94 & 1.24 & 7.12 & 97.23 & -6.87 & CARN & 200 \\
			16544 & 8.30 & 7.09 & G2V & 19.13 & 43.98 & 1 & 0.94 & 4.75 & 95.67 & -5.77 & BPMG & 24 \\
			17338 & 8.69 & 7.20 & G8V & 17.57 & 50.75 & 1 & 1.03 & 5.48 & 87.87 & -5.99 & OCT & 35 \\
			30034 & 8.60 & 6.98 & K1V & 19.94 & 65.45 & 0.86 & 1.04 & 6.13 & 99.73 & -5.99 & CAR & 45 \\
			30314 & 6.15 & 5.04 & G1V & 41.89 & 12.52 & 1.08 & 1.07 & 1.40 & 209.44 & -7.32 & ABDMG & 149 \\
			50032 & 8.58 & 7.06 & K2V & 24.33 & 27.26 & 0.90 & 0.8 & 2.69 & 121.65 & -5.49 & UCL & 16 \\
			53837 & 7.16 & 6.05 & G2V & 30.03 & 15.46 & 1.08 & 0.95 & 1.73 & 150.16 & -6.14 & LCC & 15 \\
			78264 & 8.19 & 7.21 & F7V & 14.43 & 50.50 & 1.16 & 1.1 & 5.82 & 72.16 & -5.83 & UCL & 16 \\
			81295 & 9.35 & 8.24 & G3/5V & 11.09 & 101.70 & 1.06 & 0.94 & 11.29 & 55.47 & -4.89 & UCL & 16 \\
			\hline
		\end{tabular}
	\end{center}
	\caption{10 out of 38 identified targets for the kernel-nuller discovery observing program, derived from the proper motion anomaly (PMa, \cite{2022Kervella}) catalogue of Hipparcos and Gaia EDR3 astrometry. The table includes the stellar R \& K magnitudes, spectral type, parallax ($\varpi$), $\Delta v_\mathrm{T,G3}$ is the norm of the tangential PMa vector converted to linear velocity using the Gaia EDR3 parallax, $m_{\mathrm{\star}}$ the mass of the primary star, $R_\star$ the stellar radius and $m_{\mathrm{P}}$ is the mass of a companion assuming a 5 AU orbit. The angular separation estimate is given by the ratio between the semi-major axis (assumed to be 5 AU) and the distance to the system. The estimated contrast $c$ of the companion is given as $\log(c)$. The full names of the young stellar associations listed are AB Doradus (ABDMG), $\beta$ Pictoris ($\beta$PMG), Carina (CAR), Carina-Near (CARN), Lower Centaurus Crux (LCC), Octans (OCT), Tucana-Horologium Association (THA) and Upper Centaurus Lupus (UCL). The ages of the young associations are from \cite{2018Gagne}.}
	\label{table:1}
\end{table*}

   \begin{figure}
   \centering
   \includegraphics[width=0.35\textwidth]{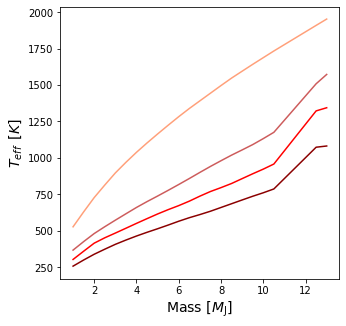}
   \includegraphics[width=0.35\textwidth]{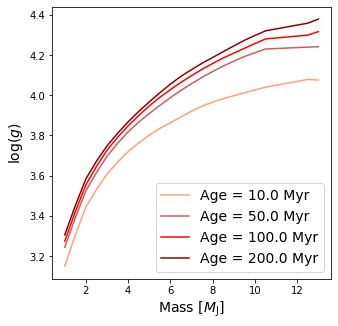}
      \caption{AMES-Cond isochrones computed with a cloudless atmosphere as a boundary condition of the interior structure. \emph{Top Panel:\/} Evolution of the effective temperature $T_{eff}$ of giant planetary companions as a function of mass for different ages. \emph{Bottom Panel:\/} Evolution of the surface gravity $\log(g)$ of giant planetary companions as a function of mass for different ages. The surface gravity $g$ is in cgs units.}
         \label{isochrone}
   \end{figure}

    Over the past decade, the majority of directly imaged exoplanets have been discovered from "blind" surveys where targets are selected based on age and distance from Earth. These detections have mostly been in the near-infrared using 8-m ground-based telescopes coupled with extreme adaptive optics (AO) and coronagraphy. The most recent blind direct imaging surveys such as Gemini-GPIES \citep{2019Nielsen} and SPHERE-SHINE \citep{2017Chauvin} have shed light on the occurrence rates of giant planets between 10 - 100 AU and the formation mechanisms involved. However, these surveys have yielded few planet detections, for instance, the GPIES survey detected 9 companions from the $\sim$300 stars imaged. The planetary mass companions around three of the imaged stars (HR 8799; \cite{2008Marois}, $\beta$ Pic; \cite{2009Lagrange}, HD 95086; \cite{2013Rameau}) were already known prior to the survey. Young stars ($<$ 100 Myr) have been natural targets for both the GPIES and SHINE surveys as their young age corresponds to bright young exoplanets thus reducing the contrast ratio required for successful observations. The close distance of these surveys ($\sim$ 100 pc) should allow for exoplanet detections at closer separations to their host stars, although the exoplanets detected have been at wide separations ($>$ 20 AU) typically because the Gemini and SPHERE imaging systems are limited to observing beyond an IWA of a few $\lambda /d$ (92 AU at $\sim$ 111 pc). Indirect observation methods such as radial velocity and primary transit are more sensitive at close ($<$ 10 AU) separations, though here the presence of the planets is inferred whereas direct imaging provides the photons from the planets themselves. The angular resolution provided by long-baseline interferometry and an on-axis high-contrast imaging mode will enable the direct detection of planetary companions in a region of the parameter space that has thus far only been covered by indirect detection methods.
    
    \subsection{Combining astrometry and interferometry}
    \label{sec:astro_direct}
    
    To be efficient, an interferometric direct detection survey of extrasolar companions should target objects that are already suspected of harbouring one or more low-mass companions. With the aid of the unprecedented accuracy of Gaia \citep{2016GaiaCollaboration} astrometric measurements and its sensitivity to sub-stellar companions, new techniques such as the proper motion anomaly (PMa, \cite{2019Kervella,2022Kervella}), the $\delta \mu$ method \citep{2019Fontanive,2020Bonavita,2022Bonavita} and astrometric accelerations \citep{2018Brandt,2019Brandt,2021Brandt} demonstrate how dynamical evidence which previously only gave access to stellar companions can now provide access to sub-stellar companions. These techniques are all based on changes in stellar proper motions measured between two different astrometric catalogues covering long time baselines. The proper motion values measured between the catalogues should be identical for an isolated star. However, the tug of a companion will cause the position of the star to deviate slightly from what would be expected if it was isolated. This will result in a difference in proper motion measurements between the catalogues and can thus be interpreted as an indication of the presence of an unseen companion around a single star.
     \smallbreak
    The COPAINS Survey \citep{2019Fontanive,2022Bonavita} conducted with SPHERE/VLT is a recent successful use case of this $\delta \mu$ method, that led to the detection of 10 companions (6 of stellar mass and 4 brown dwarfs) out of a 25 target sample. In this section we consider such a targeted approach for a discovery observing program with a kernel-nulling recombiner at the focus of the VLTI.
    
    \subsection{Target selection}
    \label{sec:PMa_catalog}
    
    We use the PMa catalogue \citep{2022Kervella} of Hipparcos and Gaia EDR3 astrometry for $\sim$11,000 stars as a proxy for giant planetary companions (1 - 13$M_{\mathrm{Jup}}$). Since the VLTI is located in the southern hemisphere we only consider targets from the PMa catalogue with dec $< 16^\circ$. The sensitivity of the PMa technique to sub-stellar companions decreases with the distance to the target hence we restrict our target list to stars located within 100 pc ($\varpi>$10 mas) of the Sun. The sensitivity of the PMa technique will also decrease as the mass of the primary star $m_{\mathrm{\star}}$ increases. We therefore impose a conservative range of 0.8\(M_\odot\)$< m_{\mathrm{\star}} <$ 1.2\(M_\odot\) on the primary mass which has been estimated via various references in \cite{2019Kervella}. As an indicator of binarity, we follow the approach of \cite{2022Kervella} and consider as probable companion hosting stars all the targets with an EDR3 PMa S/N $>$ 3. Assuming a circular orbit of radius $r$ for the companion, its mass $m_{\mathrm{P}}$ is related to the PMa signal $\Delta\mu$ by the following expression shown in \cite{2019Kervella} as:

    \begin{equation} \frac{m_{\mathrm{P}}}{\sqrt{r}} = \sqrt{\frac{m_{\mathrm{\star}}}{G}} v_{\mathrm{1}} = \sqrt{\frac{m_{\mathrm{\star}}}{G}} \left( \frac{\Delta\mu}{\varpi} \times 4740.47\right), \label{PMa} \end{equation}

    \noindent where $G$ is the gravitational constant, $v_{\mathrm{1}}$ the tangential orbital velocity of the primary and $\varpi$ its parallax. The $m_{\mathrm{P}}$ values are estimated in the PMa catalogue at an orbital separation of 5 AU. In this scenario, we select the targets for which the astrometric signature is compatible with the presence of giant planetary mass companions. We then apply the Banyan $\Sigma$ algorithm \citep{2018Gagne} to this list of targets in order to determine their membership probability to young stellar associations. The Bayesian probability reported by Banyan $\Sigma$ is based on the stellar coordinates, proper motion, radial velocity and distance of the targets. 38 targets are identified to have a probability $>$ 95$\%$ of belonging to a young association and we show a sample of these in Table~\ref{table:1}.
    \smallbreak
    Now that we have an estimate on the age of the giant planetary companions potentially orbiting these targets, we can use isochrones from an evolutionary model to provide us with the photometric properties of the planet as a function of age. To do this we used $species$ \citep{2020Stolker}, which is a toolkit developed for the spectral and photometric analysis of directly imaged exoplanets. Several evolutionary models are supported by $species$ and for this analysis we use the AMES-Cond model. The AMES-Cond isochrones shown in Fig.~\ref{isochrone} assume a hot start (i.e. high initial entropy) for the planet and one can use them to estimate its temperature $T_P$ and surface gravity $g$. With these parameters we can estimate the L-band planet/star flux ratio for self-luminous planets assumed to have a blackbody spectrum as:
    
    \begin{equation} c = \left( \frac{R_P}{R_\star} \right)^2\frac{T_P}{T_\star}, \label{Cont} \end{equation}
    
     \noindent where $R_P$ is the planetary radius that is derived from $g$, $R_\star$ the stellar radius (see Table~\ref{table:1}) and $T_\star$ the star's effective temperature that is approximated from its spectral type. Fig.~\ref{cont_sep} shows how the 38 identified companion candidates are distributed in an angular separation - contrast plot. To achieve a spectral resolution R$\sim50$, the analysis by \citet{2020Ireland} shows that reaching a contrast better than $c = 10^{-5}$ over a 1-hour integration with the ATs requires the targets to be brighter than $M=9$ in the science band-pass (here the L-band). Moreover, the ability to observe the different targets is also dependent on the performance of the upstream AO system NAOMI \citep{2019Woillez} which has a limiting magnitude of R = 12. We cross-matched our sample of targets with the USNO-B catalogue \citep{2003Monet} to find the corresponding R-band magnitudes. Thus we can confirm that all our 38 targets are bright enough to be observed by the VLTI ATs and lead to sensitive contrast detection limits. Table~\ref{table:1} lists the properties of 10 out of the 38 identified targets for the kernel-nuller discovery observing program.
     
    \subsection{Kernel-nulling at the VLTI}
    \label{sec:exo_vlti}

\begin{figure}
   \centering
   \includegraphics[scale=0.41]{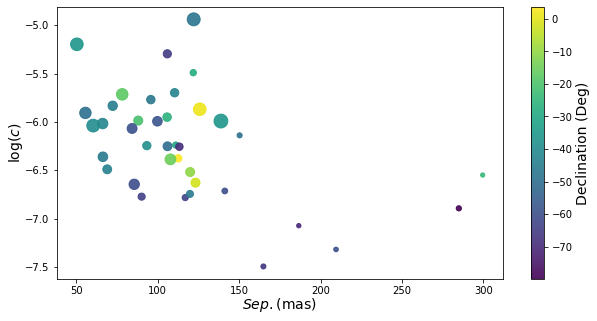}
   \caption{Identified companion candidates to the targets identified for the kernel-nuller discovery observing program. These 38 targets are derived from the proper motion anomaly (PMa) catalogue \citep{2022Kervella} of Hipparcos and Gaia EDR3 astrometry. The companion contrast $c$ is presented as a function of the angular separation which is given by the ratio between the semi-major axis (assumed to be 5 AU) and the distance to the system. The colour of the markers indicates the declination angle of the target system and the size depicts the mass of the companion ($M_{\mathrm{Jup}}$).}
              \label{cont_sep}%
    \end{figure}

\begin{figure}
   \centering
   \includegraphics[scale=0.6]{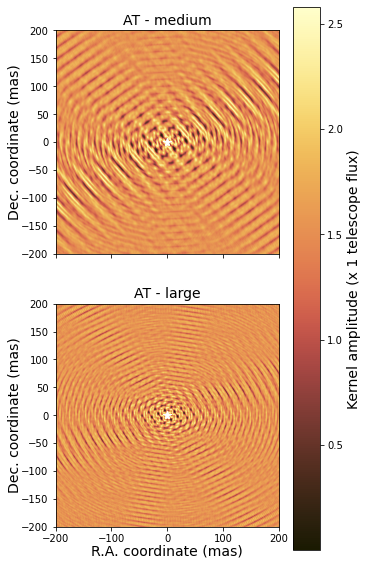}
   \caption{Kernel global throughput maps of a kernel-nuller at the focus of the four VLTI ATs, observing for 1 hour the $\pm$200 mas field of view surrounding the target HIP 116748. \emph{Top Panel}: The throughput map computed with the medium (K0-G2-D0-J3) AT configuration. \emph{Bottom Panel}: The throughput map computed with the large (A0-G1-J2-J3) AT configuration. Both maps share the same colour scale where the transmission is expressed in units of the kernel amplitude collected by one telescope. A white star in each map marks the location of the central star where the rejection by the nuller is optimal.}
        \label{mag_map}%
    \end{figure}

   Considering that telescope time is limited, using the UTs for kernel-nulling observations would only be warranted in the context of an exoplanet characterisation program. We have seen from the colinearity maps shown in section \ref{sec:colin} that the medium and large AT configurations are both capable of providing the spatial resolution needed for exoplanet detection on account of their long baselines. To request an imaging program with the VLTI ATs, ESO\footnote{\url{https://www.eso.org/sci/facilities/paranal/telescopes/vlti/configuration/P111.html}} requires a user to select a minimum of two configurations. Thus, we opt for the medium and large AT quadruplets for our discovery observing.  
    \smallbreak
    The single-mode (diffraction limited) field of view of the ATs for L-band observations extends to $\sim 400$ mas. This outer working angle (OWA) adequately covers the expected range of angular separation for the 38 companion candidates represented in Fig.~\ref{cont_sep}. The large AT quadruplet configuration features a maximum baseline length of 132.4-m which sets an IWA of 5.4 mas (6.9 mas for the medium). The IWA and OWA constraints suggest that the kernel-nuller can detect giant planets on orbits ranging from 0.24 to 18 AU (0.31 - 18 AU for the medium) for an object located at a distance of $\sim$ 45 pc (median distance of identified companion candidates).
    \smallbreak
    An observer in charge of the planning of this program can use the kernel global throughput map we've presented in section \ref{sec:kmag} to predict the probability of detection at the expected location of a potential companion. Fig.~\ref{mag_map} shows the maps computed with the medium and large AT configurations for one of the identified targets - HIP 116748 ($c\sim10^{-6}$). A companion orbiting at 5 AU from this host star would be located at an angular separation of $\sim$ 113 mas. Over the course of a 1-hour observation, the throughput maps go through a $\sim10^\circ$ rotation. The average throughput of the nuller across the field of view reaches $\sim$3: the planet flux collected by three out of the four sub-apertures contributes to the detection of $\sim 2,300$ photons (assuming an overall efficiency of $\sim3.2\%$ as reported by \cite{2022Laugier}). The corresponding average magnitude of the kernel-null (see Fig.~\ref{mag_map}) is $\sim1.55$. Both configurations provide a good detection completeness at the estimated location of the companion. For targets at higher (less favourable) declinations ($> -30^\circ$), the large configuration should be preferred.
	\smallbreak
    Fig.~\ref{cont_sep} shows that the average expected contrast ratio for the identified companion candidates is $c\sim10^{-6}$. According to \citet{2020Ireland}, reaching this level of contrast puts some very stringent requirements on the quality of the fringe tracking ($\sim3$ nm RMS) and the stability of the intensity fluctuations (0.01 \% RMS). These requirements are certainly challenging but achieving them would enable key science goals such as the detection of young giant exoplanets at separations $<$ 10 AU and the atmospheric characterisation of their atmospheres with a moderate spectral resolution (R$\sim50$). A detailed analysis of the systematic errors introduced by the non-ideal nature of a nuller is beyond the scope of this paper. However, this can be the subject of a future publication as it is required to draw more definitive conclusions on the sensitivity limits of our approach.
    \smallbreak
    The infrastructure of the VLTI is currently on an aggressive upgrade path: moving from the first generation FINITO fringe tracker \citep{2008LeBouquin} to that of GRAVITY \citep{2019Lacour} has successfully reduced the fringe tracking residuals on the ATs from $\sim150$ to $\sim100$ nm RMS, and other upgrades of the VLTI infrastructure are on their way. High-contrast detection of planetary companions using nulling is the objective of the NOTT instrument \citep{2022Laugier}: one of the modules of the planned Asgard instrument suite to be installed at the focus of the VLTI in 2025. The Asgard suite also includes HEIMDALLR \citep{2018Ireland}: a high precision fringe tracker and injection stabilisation module. A HEIMDALLR prototype \citep{2023Cvetojevic} was tested in the laboratory and demonstrated that for targets brighter than $K = 11$, closed-loop fringe tracking residuals can be brought down to $\sim4$ nm RMS. To assess ultimate sensitivity, further analysis of the longitudinal dispersion between the K- and the L-band will be required. Assuming that the Asgard shared infrastructure fully addresses these effects and the operational challenges of non-common path aberrations, this ambitious program remains achievable with the VLTI.

    \section{Conclusions}
    
    The analysis performed in this paper shows that by taking into account baseline projection effects which are induced by Earth rotation, we introduce some diversity in the response of a nuller as a function of time. This response is depicted by transmission maps, which we have shown for both the raw nuller outputs and their subsequent kernel outputs - which are the primary intended observable quantities of the kernel-nuller. To further comment on the properties of the kernel-nuller, we examine the absolute value of the kernel output and present maps of this absolute value which we call kernel magnitude maps. With enough evolution of the response of the nuller, we show that the so called 'blind spots' that we see in the snapshot throughput maps do eventually fill in.
    \smallbreak
    The colinearity maps we have presented show that the increased diversity in the response results in a predictable modulation of the output of the nuller if a potential off-axis companion is present at a specific location. These maps are different for the various four telescope arrays offered at the VLTI on account of their varying baseline lengths. The location of the maximum of colinearity in these maps can be used as a first estimate for the position ($\alpha$, $\delta$) and contrast $c$ of the actual companion, which can then serve as the starting point in a parametric $\chi^2$ minimisation procedure. While we have been able to infer the position of a single companion with the colinearity map, this investigation could be extended to multiple companions in the FOV. For such an arbitrary high contrast scene, the image can be reconstructed using a variety of inverse model techniques such as SQUEEZE \citep{2010Baron} or MACIM \citep{2006Ireland}.
    \smallbreak
    We have assessed the characteristics of a kernel-nuller discovery observing program for the direct imaging of exoplanets. By exploiting the synergy between astrometry and direct imaging we show the target selection process for this program. The complete analysis of the targets identified is beyond the scope of this paper as one would also need to take into account other indicators of binarity such as the renormalised unit weight error (RUWE; \cite{2021Lindegren}) and determine if the potential giant planetary mass companions identified are gravitationally bound to their host star. Additionally, the possible positions of the companion generating the astrometric signature can be further limited by using $\Delta v_\mathrm{T,G3}$ as shown by \cite{2022Bonavita}. Furthermore, with all subsequent GAIA data releases, the sensitivity of the PMa technique will continue to improve.
    
    \begin{acknowledgements}
    This project has received funding from the European Research Council (ERC) under the European Union’s Horizon 2020 research and innovation program (grant agreement CoG - 683029). DD acknowledges the support from the ERC under the European Union's Horizon 2020 research and innovation program (grant agreement CoG - 866070). 
    \end{acknowledgements}

%
\bibliographystyle{aa} 
\bibliography{ref.bib} 

%

\end{document}